\documentclass[preprint]{aastex}

\begin{document}

\slugcomment{Not to appear in Nonlearned J., 45.}

\shorttitle{Near-infrared EBL}

\shortauthors{Matsumoto et al.}

\title{IRTS observations of the near-infrared extragalactic background light}

\author{T. Matsumoto, 
  S. Matsuura,
  H. Murakami,
  M. Tanaka\altaffilmark{1},
  M. Freund\altaffilmark{2}
  and
  M. Lim\altaffilmark{2}}
\affil{Institute of Space and Astronautical Science,
  Japan Aerospace Exploration Agency,
  Kanagawa 229-8510, Japan}

\author{M. Cohen}
\affil{ Radio Astronomy Laboratory, University of California, Berkeley, 
      California CA 94720, U.S.A. }

\author{M. Kawada}
\affil{Department of Physics, Nagoya University, Nagoya 464-8602, Japan}

\author{M. Noda}
\affil{Nagoya Science Museum, Nagoya 453-0037, Japan }

\altaffiltext{1}{Present address is
National Astronomical Observatory of Japan,
Mitaka 181-8588, Japan}
\altaffiltext{2}{Now out of ISAS/JAXA}


\begin{abstract}
We have searched for near-infrared extragalactic background light (EBL) in the 
data from the Near-InfraRed Spectrometer (NIRS) on the Infrared Telescope 
in Space (IRTS). After subtracting the contribution of faint stars and the 
zodiacal component based on modeling, a significant  isotropic
emission  is obtained in the wavelength bands from 1.4 $\mu$m to 4.0 $\mu$m.
The spectrum is stellar-like, but shows a spectral jump from the optical EBL.
The emission obtained is isotropic over the observed sky, 
and the in-band flux amounts to $\sim 35$ nWm$^{-2}$sr$^{-1}$
which is too bright to be explained by the integrated light from faint galaxies. 

Analyses of COBE/DIRBE data, after removal of starlight, show essentially the 
same result within the uncertainty in the zodiacal light model, which implies that
the isotropic emission observed by IRTS/NIRS is of extragalactic origin.  

Significant fluctuations in sky brightness were also detected which cannot be 
explained by fluctuations due to faint stars, zodiacal components and normal galaxies. 
The excess fluctuation amounts to $ \sim 1/4 $ of the excess emission over 
the integrated light of galaxies and is consistent with
fluctuations observed by COBE/DIRBE. A two-point correlation analysis shows
that IRTS/NIRS data has an angular scale of fluctuations of a few degrees.

The spectrum and brightness of the observed excess EBL emission could
be explained by the
redshifted UV radiation from the first generation of massive stars
 (Population III stars) that caused the reionization
of the Universe. Recent WMAP observations of
the CMB polarization have indicated that reionization occurred at $ z\sim 17 $ 
or earlier, while the spectral jump around 1 ${\mu}$m in the observed excess EBL
suggests that the Pop.III star formation terminated at $ z \sim 9 $.
The observed fluctuations, however, are considerably larger than the theoretical predictions for the Pop.III stars.

\end{abstract}


\section{Introduction}

The infrared extragalactic background light (infrared EBL) provides
an important clue to our understanding of the early universe 
and the evolution of galaxies.
The near-infrared ($1 \sim 5$ $\mu$m), and 
far-infrared/submillimeter (50 $\mu$m $\sim$ 1 mm) wavelength bands are regarded as useful 
windows for detecting the EBL since the sky brightness is very low there. 

It has been assumed that redshifted starlight from high redshift
galaxies constitutes the near infrared EBL, which
must be  important to understanding the energy generation 
during the galaxy formation era. The fluctuation of the near infrared EBL 
is also crucial to understanding the evolution of the large scale structure of the Universe.
Much effort has been devoted to the detection of the near infrared EBL from the ground 
but without success
because of bright atmospheric and instrumental emissions. 
Matsumoto et al. (1988) first attempted a rocket experiment and 
succeeded in observing the near-infrared sky brightness.
Several more rocket experiments were attempted (Noda et al. 1992, Matsuura et al. 
1994). However, they had great difficulty in subtracting the foreground emission
because of the limited observing time. 
Recent space observations from the COsmic Background Explorer (COBE) 
and the InfraRed Telescope in Space (IRTS) have enabled the first quantitative 
analysis of the infrared EBL ( see the review by Hauser and Dwek 2001,
and references therein).

The Diffuse InfraRed Background Explorer (DIRBE) instrument on COBE was designed
for observations of the infrared EBL. However, observation of the near infrared EBL with
DIRBE was difficult due to the large beam size of $0.7^{\circ}$ that led to serious confusion 
from unresolved foreground stars. Several authors, however, attempted to subtract
the contribution of foreground stars and obtained a significant isotropic
emission in the J, K, and L bands that was too bright to be explained by the 
integrated light of faint galaxies (Dwek \& Arendt 1998, Gorjian et al. 2000,
Wright \& Reese 2000, Wright 2001, Cambr\'{e}sy et al. 2001). 
The detection of the fluctuation of near infrared EBL in the COBE star-subtracted 
data was also reported by Kashlinsky and Odenwald (2000), although 
not with high significance.

On the other hand, the Near InfraRed Spectrometer (NIRS) on  
IRTS had the dual advantages of a narrow beam size
and spectroscopic capability, even though the sky surveyed was 
limited (Noda et al. 1996).

In this paper, we present the IRTS observations and the result of
data analyses to search for the near infrared EBL in the IRTS/NIRS data (section 2). 
We then present the residual isotropic emission in wavelength bands from 
1.4 $\mu$m to 4.0 $\mu$m (sections 3) and also fluctuations of the sky 
brightness (section 4) that cannot be explained by known sources of foreground 
emission. We also discuss its cosmological implications based on 
recent WMAP results and show the future prospects for space 
observations (section 5). The essential results are similar to those 
reported in conference proceedings (Matsumoto et al. 2000, Matsumoto 2001).

\section{ Observation and selection of the data }

NIRS is one of the focal plane instruments of IRTS, and was optimized to obtain spectra 
of the diffuse background (Noda et al. 1994). NIRS covers the wavelength 
range from 1.4 $\mu$m to 4.0 $\mu$m, with a spectral resolution of 0.13 $\mu$m,
in 24 independent bands. 
The beam size is 8 arcmin square which is considerably smaller than that of DIRBE.

IRTS was one of the mission experiments on the small space platform, SFU, that 
was launched on March 18, 1995. The orbit was a low-inclination near-earth 
orbit and IRTS surveyed the sky avoiding the sun and the earth. The IRTS observations lasted for about 30 days, 
during which 7\% of the sky was surveyed (Murakami et al. 1996). 

During the entire IRTS mission, NIRS worked as expected. Details of the 
flight performance of NIRS can be found in Noda et al. (1996), and we 
concentrate on topics related to the absolute background measurement in this paper.

Fig.1 shows a typical output signal of the detector which had an operation
cycle of 65.54 sec. Detectors were of the charge-integrating type, and a
typical ramp curve is shown in the upper panel of Fig.1. 
The dark levels were defined as the signals when the cold shutter was
closed for 8.192 sec every 65.54 sec. The integrated charges were
discharged (reset) every 65.54 sec, the same period but a half cycle shifted 
from the operation of the cold shutter. The calibration lamp was lit every 17.5
min to monitor detector response.
The lower panel of Fig.1 indicates the differentiated signal in which the stars 
detected are designated by 5-pointed symbols.

An anomalous change of the DC level was seen just after the reset, 
with an exponential decay 
with a time constant of $ \sim 0.3 $ sec for all bands.
To avoid the influence of this anomaly, the data for 4 sec following the
reset were not used in this analysis.

Before the aperture lid opened,
no significant signal over the dark levels was detected, which ensures there was 
no internal emission in the IRTS cryostat. 
Fig.1, however, clearly shows a significant signal was 
detected when the shutter was opened in orbit. The detector responsivities 
derived from the signals of the calibration lamp were stable except 
for the periods when the SFU passed through the South Atlantic Anomaly (SAA). 

No significant change in the signals both from the sky or in dark 
levels was found between day and night in one orbital revolution. 
Furthermore, no dependence of detector signals on  
elevation angle from the Earth's
surface was detected. This implies that the baffling system 
worked fairly well and the detectors did not suffer from stray
light due to radiation from the Sun and Earth. 

During the flight, many bright stars were observed, and
a small amount of light leakage through the closed shutter was observed 
in the short wavelength bands. These amounts were measured during flight
as a small correlation between the dark levels and the sky brightness,
but the correction to the observed signals could be made fairly well. 
The largest leakage, found at 1.43 $\mu$m, was 
$3.9\pm0.3\%$ but no leakage was 
detected for wavelength bands longer than 2.0 $\mu$m. 

In the initial phase of observations, NIRS observed the environmental OH 
emission in the Southern hemisphere, which had a peak at about 2.8 $\mu$m. 
This emission exponentially decayed with a time constant of 5 days, and
disappeared by the latter half of the IRTS mission.
Passages through the SAA and the entry of lunar 
radiation into the telescope tube also seriously affected the detectors. 
To avoid these effects, only data from the last 5 days, on orbits that did not 
pass through the SAA, did not suffer from lunar radiation, and were in
the Northern sky, were used for this analysis.

Absolute calibration was satisfactorily achieved with several standard stars
observed during the flight (Murakami et al. 2003).
The magnitude system for the NIRS bands was constructed 
so that the observed spectra
of the standard stars were consistent with previous observations (Cohen et al. 1999).  
The beam pattern was also measured by using the passage of bright stars,
and the corresponding conversion factors to sky brightness were obtained. 

To reduce the contribution from faint stars, the sky at high 
galactic latitudes ($ b>40^{\circ}$) was chosen. The highest galactic latitude was
$58^{\circ}$, while the ecliptic latitude ranged from $12^{\circ}$ to $71^{\circ}$ in 
the selected data set.

Data were taken within a 5-sec integration during which no distinguishable stars 
and no cosmic ray hits were detected in any NIRS
bands. After subtracting the dark current when the cold shutter was closed, 
signals from the sky were obtained. 

During 5-sec, the telescope axis moves about 20 arcmin along a great circle, which
results in a trapezoidal beam pattern as shown in Fig.2. 
The effective beam size for diffuse extended sources was 8 arcmin $\times$ 20 arcmin.  
However, the contribution from point sources depended on their position in the beam.

Celestial coordinates of the data points were determined by use of the focal plane star sensor
and other attitude sensors. The error in the attitude determination for the data set 
was estimated to be about half a beam size (${\sim 4 }$ arcmin).

Finally, complete spectra of the sky were secured at 1010 positions which were free from 
contamination and cosmic rays. 

Fig.3 indicates the positions of the spectra acquired in galactic coordinates. 
Ecliptic coordinates are also shown by dashed lines. The sun lay close to the
Spring equinox, and solar elongation angles were 
$\sim 90^{\circ}$ during the observations.

Fig.4 shows our averaged spectrum of the sky at high ecliptic latitude
($ \beta>70^{\circ}$) in which the data of the DIRBE darkest sky (Hauser et al. 1998) 
are also presented. The surface brightness is expressed as
$ \lambda \cdot I_{\lambda}= \nu \cdot I_{\nu} $, in units of 
nWm$^{-2}$sr$^{-1}$ which is used throughout this paper.\\
Fig.4 implies that 
both the NIRS and DIRBE data are fairly consistent despite 
the difference in regions observed and beam sizes.

\section{Subtraction of foreground emission}
\subsection{Contribution of faint stars}

One of the superior characteristics of NIRS is that fainter stars 
could be resolved and removed due to the reduction in confusion owing to the 
smaller beam size than COBE/DIRBE.

The first step toward subtracting the stellar component (star light) is to 
establish the number density of 
stars (i.e. the $log N / log S$ relation). Fortunately, NIRS detected several tens of 
thousands stars during the IRTS mission, and a complete catalog is open
to the public (Yamamura et al. 2003).
Although this catalog is rather shallow (completeness limit at 2.24 $\mu$m is 
$\sim$ 7.5 mag), we have been able to predict the observed 
$log N / log S$ relations (within the Poisson error bars) down to the completeness 
limits for all 24 NIRS bands by using Version 5 of the  ``SKY" model for the point source sky 
of the Galaxy (Cohen 1997).
The success of these predictions justifies our extrapolation of the 
observed NIRS star counts to 30th mag using SKY.

The second step is to determine the contribution of stars that are too faint to be 
resolved in the NIRS data stream. We first define the cut-off magnitude in a given
wavelength band as the flux 
that corresponds to the noise having the same spatial frequency as the beam 
pattern. This means that all stars brighter than the cut-off magnitude in any 
bands are removed from the acquired data.
For example, the cut-off magnitude obtained in the 2.24~$\mu$m band
is 10.45 mag. Uncertainties for the cut-off magnitudes are $\pm 0.5$ mag in all NIRS bands.
It must be noted that the cut-off magnitude is identical for the whole
of the trapezoidal beam pattern (8 arcmin $\times$ 26 arcmin, Fig.2) owing to the procedure
for data acquisition.

With the $log N / log S$ relations predicted by the SKY model, and using these cut-off magnitudes, 
we calculated the total surface brightness due to stars between 30th and the cut-off
magnitudes at $ b=42^{\circ}, l=80^{\circ}$, $ b=45^{\circ}, l=85^{\circ}$, 
and $ b=48^{\circ}, l=90^{\circ} $ along the scan path.
The calculation was accomplished by a Monte Carlo method applying the $log N / log S$ relations
for the beam pattern shown in Fig.2. In total 100,000 trials were performed for each of the three
galactic latitudes, 24 wavelength bands, and three cut-off magnitudes. Fig.5 shows the
average surface brightness thus obtained at 1.63, 2.14 and 3.48 $ \mu$m as 
a function of $ cosec(b) $. Since excellent linearity is found, we have
estimated the stellar contribution at any point assuming the $ cosec(b) $ law.
The dependence on galactic longitude was neglected since stars at high
galactic latitudes are mostly of local origin and are not affected by galactic
structure. This procedure is justified later by our own data. 
The contribution from faint stars in the 2.14-$\mu$m band is about $10\%$ 
of the observed sky brightness at high ecliptic latitude.

There are two kinds of uncertainty in the contribution of faint stars.
The first is that due to the error in the cut-off magnitudes ($ \pm 0.5$ mag).
This error can be estimated rather easily with the SKY model (see Fig.5) and are taken 
into consideration in our estimation of the total error. 

The second uncertainty is the model itself. Typical uncertainties in SKY's predicted values of 
$log N$ are $30 \sim 40\%$ at the faintest relevant magnitudes 
(e.g. Minezaki et al. 1998, their Fig.2). However, the diminishing contribution 
of ever fainter sources to the predicted surface brightness leads to much smaller
uncertainties in sky brightness, formally quantified by Cohen (2000) as 
$\sim\pm9\%$ in a typical application to surface brightness at 3.5 $\mu$m.

In order to verify the $log N / log S$ relations directly, we used 2MASS
data. The direct subtraction of 2MASS stars from the observed data is unfortunately 
difficult due to the poor attitude determination. However, the 2MASS catalog
is useful enough to validate the SKY model. 
2MASS stars fainter than the cut-off magnitudes are picked up
by the IRTS data points at $ b=42\pm1^{\circ}$, 
$ b=45\pm1^{\circ}$ and $ b=48\pm1^{\circ}$ in the 
H (1.66 $\mu$m) and K (2.16 $\mu$m) 
bands and converted into surface brightness. In this analysis, a flat top beam 
(8 arcmin $\times$ 20 arcmin) was used for simplicity. 
An estimation using the SKY model was attained analytically for a flat top beam,
and it was found that the trapezoidal beam pattern (Fig.2) provides 
7 and 12 \% less contribution to the surface brightness and to its fluctuation than
the flat top beam pattern, respectively.
Since the 2MASS catalogue has completeness limits of 15.3 mag  for the
H band and 14.3 mag for the K band, the contributions of stars fainter than the 
completeness limits are supplemented by using the model. Their contribution to
the surface brightness is about 20 \%.
In Fig.6, the surface brightness obtained from 2MASS stars is compared with 
that of the SKY model 
for the 1.63 and 2.14 $\mu$m bands (which are very close to the H and K bands 
of 2MASS catalog), respectively. The results of the 2MASS analysis are fairly consistent 
with the model and we assigned an error to the model as $ \pm  8 \% $ 
for all wavelength bands, adopting the worst case in Fig.6.

The integrated light of the 
faint stars extrapolated by the SKY model was subtracted from the observed 
sky brightness in each wavelength band. 
This ``star subtracted data'' is used for further analysis.

\subsection{Subtraction of the zodiacal component}

Kelsall et al. (1998) constructed a physical model of the interplanetary 
dust (IPD) using the seasonal variation of the zodiacal light and 
thermal re-emission. Based on this model, we calculated the brightness of 
the zodiacal component corresponding to our wavelength bands and 
observed points in the sky. The wavelength dependent 
parameters (phase function parameters, albedos, emissivity modification
factors) in the Kelsall model are obtained by interpolation between those of the COBE
wavelength bands. The characteristic feature of the resulting NIRS  
zodiacal model is that
there exists excellent linear proportionality between
NIRS wavelength bands, although slight deviations are found at the longest wavelengths. 
In other words, the spectral shape of the zodiacal component is essentially the same
for the regions observed by NIRS. This can be understood from the fact 
that solar elongation angle for the NIRS data points 
is fairly constant ($ \sim 90^{\circ}$),
and the dust in the smooth cloud makes the dominant contribution to the zodiacal
component. 
    
We tried to subtract this modeled brightness from the ``star-subtracted-data''. 
First of all, we undertook a correlation analysis between ``star-subtracted-data'' 
and modeled zodiacal components. Fig.7 shows the correlation diagrams 
at (a) 1.63 $\mu$m, (b) 2.14 $\mu$m and 
(c) 3.48 $\mu$m, respectively. The correlation is excellent for all bands.
As a minor correction 
to the model, we applied linear fits to the correlation diagrams 
leaving the slopes as the free parameters. 
The slopes are indicated in Fig.8 as a function of wavelength in which
the longest wavelength bands have a little steeper than 1.0. 
We regard that these corrections to the slopes as due to 
differences of calibration between COBE and IRTS, and 
uncertainty in interpolating the optical parameters in the model. This
can be justified by the excellent linear proportionality between NIRS
wavelength bands. This procedure can be understood as an
iterative process to improve the NIRS zodi model so that the residual emission
is as isotropic as possible over the sky observed by IRTS/NIRS. 
The correction factors for the K and L bands, however, 
are less than 10 \%, which is within the uncertainty of the original IPD model. 
Furthermore, Fig.7 clearly indicates that residual 
emission remains where the modeled zodiacal components are zero. 

Fig.9 shows the ecliptic latitude dependence of each star-subtracted brightness
(pluses) and modeled zodiacal component after correction (solid line) 
at (a) 1.63, (b) 2.14 and (c) 3.48 $\mu$m, respectively. 
The model brightness of the zodiacal component was
calculated for all observed positions but the results show no dependence 
on ecliptic latitude. As a result, the model brightness in Fig.9 
appears as a continuous curve.

Fig.9 clearly shows that there remains a fairly isotropic  
residual emission (crosses) after subtracting the zodiacal components
from ``star-subtracted-data''. Other wavelength bands show essentially the 
same pattern. 

\subsection{Isotropic emission}

We first investigate the existence of any galactic component in residual brightness.
Fig.10 shows the residual brightness as a function of $ cosec(b) $. 
From the top, those for 1.63 $\mu$m,
2.14 $\mu$m, 3.28 $\mu$m and 3.48 $\mu$m are plotted, respectively.
The residual brightness shows a fairly isotropic character and no clear dependence 
on $ cosec(b) $ is found in any wavelength band. This
ensures that the adopted SKY model is appropriate.

There may be near infrared diffuse emission associated with interstellar matter 
as was sought in the DIRBE data by Arendt et al. (1998). They
found no significant associated emission in the J and K bands, but detected a slight
correlation in the L band at low galactic latitude ($ b<30^{\circ}$). 
Tanaka et al. (1996) detected diffuse emission from the 3.3 $\mu$m UIR band 
on the galactic plane with the IRTS/NIRS data. This emission has a clear correlation 
with far infrared dust emission. However, detection was limited to low
galactic latitudes, $b<10^{\circ}$. This is confirmed by the fact that the residual emission 
at 3.28 $\mu$m in Fig.10 shows no clear dependence on galactic latitude.  
In summary, we conclude that the contribution of diffuse emission
associated with interstellar matter is negligible for all wavelength bands,
since our data were taken only at high galactic latitudes ($b>40^{\circ}$).

Fig.9 and Fig.10 clearly demonstrate that there remains a fairly isotropic 
emission in the IRTS data which has no dependence on either ecliptic or
galactic latitude. 
We derived this isotropic emission as the brightness at which the modeled 
zodiacal component is zero in Fig.7.
The results and errors (1 $\sigma$) are listed in Table 1.
Fitting errors are estimated statistically based on linear fits, while
errors in both star subtraction are assessed by taking into account 
uncertainties in both the cut-off magnitudes and the SKY model.
Errors in the IPD model are estimated by interpolating the uncertainties 
in the original model 
( 15, 6, 2.1, and 5.9 nWm$^{-2}$sr$^{-1}$ for the J, K, L and M bands, respectively 
(Kelsall et al. 1998)). 
Systematic errors include all instrumental uncertainties associated with 
calibration, light leakage, spectral response, etc. (Noda et al. 1996). The total error is 
obtained by summing the five kinds of uncertainty in quadrature. 
Table 1 clearly shows that the major source of uncertainty is that of the IPD model. 

Fig.11 shows the breakdown of sky brightness at high ecliptic latitude
(filled circles, as in Fig.4)
in which the zodiacal component (bars), the derived 
isotropic emission (open circles) 
and the integrated light of faint stars (open diamonds) are indicated. 
The brightness of the isotropic emission is about $\sim 1/5$ of the sky brightness.
It must be noted that the sum of the
three emission components does not reproduce
the sky brightness, because the isotropic emission is derived through statistical analysis.

Fig.12 shows our spectrum of the derived isotropic emission in comparison to
other observations. The isotropic emission component (open circles) is 
consistent with the DIRBE upper limits shown by the arrows (Hauser et al. 1998).
Open squares show the optical EBL observed by Bernstein et al. (2002).
Filled circles and filled squares indicate the results obtained for star-subtracted DIRBE 
analyses by Cambr\'{e}sy et al. (2001) and Wright and Reese (2000), respectively.
The data by Wright and Reese (2000) have been modified to use Kelsall's 
IPD model (see section 5.1).
Independent analyses show fairly consistent results, which suggests that the isotropic
emission observed by IRTS/NIRS is of extragalactic origin, that is,  EBL,  
although the sky observed by IRTS/NIRS is not extensive.
Filled diamonds show the integrated light of galaxies, that is, a 
simple integration of detected galaxies. The data in the H band are
compiled by Madau and Pozzetti (2000), and those in the 3.6 and 4.5 $\mu$m bands
are recent results from the Spitzer Space Telescope (Fazio et al. 2004).
Other estimates are by Totani et al. (2001) from the SDF (Subaru Deep Field) 
for the J and K bands, and from the HDF (Hubble Deep Field) for others. 

The solid line represents the theoretical
prediction of EBL based on observations of galaxies and using the evolutionary model of
Totani \& Yoshii (2000), whose predicted spectrum is normalized 
to $ 10$ nWm$^{-2}$sr$^{-1}$ at the top of the 
error bar at the K band.
Based on this model, Totani et al. (2001) claim that more than 90\% of the
integrated light of galaxies has already been resolved in deep galaxy counts. 
Fig.12 indicates that the near infrared EBL observed by IRTS and COBE 
is significantly brighter than the expected integrated light of faint galaxies. 
The near infrared spectrum lacks any significant feature, but there exists a 
clear jump from the optical EBL observed by Bernstein et al. (2002).
The in-band energy is $\sim 35$ nWm$^{-2}$sr$^{-1}$.

Fig.13 shows excess emission, that is, the brightness obtained by subtracting
the integrated light of galaxies (ILG) of Totani et al. (2001) (solid line of Fig.12)
from the isotropic emission. 
Optical data by Bernstein et al. (2002) are plotted without any subtraction since
the contribution of galaxies ($m_{AB}<23$ mag) has already been subtracted 
from their data. Fig.13
also indicates an abrupt spectral discontinuity around 1 $\mu$m. The solid line in Fig.13 is
the theoretical model by Salvaterra and Ferrara (2003) which will be discussed
in section 5.2.

Evidence supporting this near infrared EBL is found in intergalactic absorption 
of TeV $\gamma$ rays, that is, the inverse process of pair annihilation. 
Since the near-infrared photons cause an absorption peak around 
1 TeV, observation of the TeV $\gamma$ Blazer
at high redshift is crucial to confirm the cosmological origin of the
EBL observed by IRTS and COBE (Stecker et al. 1993, Tanihata et al. 2001). 
Recent observations of the BL Lac 
object H1426+428 at $z=0.129$ with Whipple, CAT and HEGRA show a clear
absorption feature at $\sim1$ TeV (see papers in New Astronomy Review vol.48, issue 5-6). 
This spectrum cannot be well explained by the
integrated light of normal galaxies, and requires an excess near infrared background
which has a peak around 1 $\mu$m (Aharonian et al. 2002, Mapelli et al. 2004).
There remains uncertainty in the source spectrum of the gamma rays. However, the
spectrum of the TeV $\gamma$ Blazer strongly supports an extragalactic
origin for the isotropic emission observed by IRTS/NIRS.

\placetable{tbl-1}

\section{Fluctuation analysis}

Fluctuations in the EBL are another important way 
to investigate the formation and evolution of 
galaxies and large scale structure. The effective beam size of 8 arcmin $\times$ 20 
arcmin is an adequate scale to investigate the clustering of 
galaxies at high redshift (Kashlinsky and Odenwald 2000).

For a fluctuation analysis of the NIRS data, we selected 797 data points with $ b>47^{\circ} $ 
to reduce the effect of fluctuations due to faint stars. Filled circles in Fig.14 show
the standard deviations for these data points and these are compared with
other possible sources of fluctuation.

Fluctuations due to faint stars are estimated by a Monte Carlo simulation 
for the SKY model at $ b=48^{\circ} $ which is simultaneously made when
the integrated light of faint stars is estimated in section 3.1. 
100,000 trials for each wavelength band and each cut-off magnitude are made, and
standard deviations are obtained. The result is indicated
by open diamonds in Fig.14 for which errors due to
cut-off magnitudes ($ \pm0.5$ mag) are plotted. It must be noted that the
contribution of faint stars is overestimated since the galactic latitudes of
the NIRS data used ranges from $ 47^{\circ} $ to $58^{\circ}$.

To validate the model fluctuations, we again used 2MASS stars. We picked up 
all stars between the cut-off magnitudes and completeness limits in the 2MASS catalog
for the H and K bands. Their contribution to the surface brightness
for the 797 NIRS data points was obtained. Since the 2MASS surface brightness has a dependence on
galactic latitude, we made a linear fit with $ cosec(b) $ and obtained the fluctuating
component by the fitting line. Furthermore, a correction factor of 0.88
for the flat top beam, 8 arcmin $\times$ 20 arcmin, was applied (see section 3.1).
The contribution of stars fainter than the completeness limits is neglected since
the model indicates that their contribution is negligible. The standard deviations for
this constructed  ``2MASS fluctuation data'' are shown by filled squares in Fig.14, 
and are reasonably consistent with the model estimation.

The read-out noise in each NIRS band was determined
from the noise signals when the shutter was closed. These are shown by 
open squares. 
The photon noise due to the incident radiation is shown by crosses. 
The photon noise is also overestimated, since the 
calculation is made assuming observations towards the ecliptic plane 
(Noda et al. 1996).

Fig.14 clearly shows that there 
remains a significant excess fluctuation at short wavelengths, although
read-out noise dominates at longer wavelengths. 
The excess fluctuation is obtained by subtracting the known fluctuation
in quadrature, and is shown in Fig.15 by filled circles. 
The results of the fluctuation analysis of the DIRBE data by 
Kashlinsky and Odenwald (2000) are also shown, as filled squares. 
Although their error bars are fairly large, good agreement is found with our data.

Another important feature of the observed fluctuations is the
excellent correlation between wavelength bands. 
Fig.16 shows the correlation diagram for the residual brightness at a) 1.63, 
b) 2.14 and c) 3.48 $\mu$m data to that at 1.83 $\mu$m, respectively. 
This correlation is caused by the spatial distribution of the sky
brightness and is independent of random noise (readout noise and photon noise).  
The slopes of the linear fits for these correlation diagrams indicate the colors of 
the spatially fluctuating components. Open squares in Fig.15 show the spectrum obtained
from these colors in arbitrary units. The fact that these colors are 
consistent with the spectrum of the excess fluctuation implies that the 
observed fluctuations mainly originate from
the spatial distribution of the EBL,
although a minor contribution from stellar fluctuations remains.

Open circles in Fig.15 indicate the spectrum of the excess emission (as in Fig.13)
which is very similar to that of the excess fluctuations shown by filled circles.
The fluctuation level amounts to $ \sim 1/4 $ of the excess emission.
Fig.15 implies that the observed excess fluctuations have the same origin as the 
excess emission; i.e., the excess emission fluctuates but maintains the 
same spectral shape.

Zodiacal light cannot explain such a fluctuation spectrum. 
Indeed, fluctuations of zodiacal 
emission have not yet been detected in the mid-infrared region and are less than 1 \% 
for various beam sizes (IRAS: Vrtilek and Hauser 1995, COBE: Kelsall et al. 1998, 
ISO: Abraham at al. 1998). 
The upper limits thus obtained are much smaller than the observed fluctuating 
component, that is, $\sim 1/5$ of the isotropic emission and $ \sim1/20 $ 
of the total sky brightness. 
It is unlikely that the scattering component has a much larger fluctuation 
than that of the thermal emission, so it is difficult to attribute the
excess fluctuations to any zodiacal component. The fact that the fluctuations 
of the residual emission in Fig.9 have no dependence on ecliptic latitude is 
further evidence that the observed fluctuations are not interplanetary in origin.

It is implausible to attribute the observed fluctuations to those of normal galaxies.
Fig.12 indicates that the integrated light of galaxies is $ \sim 10 $ 
nWm$^{-2}$sr$^{-1}$ at 1.4 $\mu$m which is lower than the observed
fluctuation itself. The expected fluctuation due to normal
galaxies is roughly one order of magnitude lower than the IRTS data
(Cooray et al. 2004).

The angular distribution of the fluctuations is crucial to investigate large
scale structure at high redshift. It is obvious that the typical angular scale is
larger than the COBE/DIRBE beam, $ \sim 0.7^{\circ}$.   
In order to investigate the angular scale of the fluctuations qualitatively, we
tried a two-point correlation analysis for the residual emission 
at $ b>47^{\circ} $ that was the same as used for the rms fluctuation analysis.
To secure a definite result, we defined a ``wide band brightness'' by 
integrating the brightness of the NIRS short wavelength bands from 1.43 $\mu$m to 
2.14 $\mu$m. The central wavelength is 1.7 $\mu$m, which roughly
corresponds to the H band.
For the  ``wide band brightness'', read out noise and photon noise are suppressed,
while fluctuations associated with the sky brightness must be summed, since the
sky fluctuates keeping the same spectral shape (Fig.15). 
The standard deviation of the ``wide band brightness''  is $ \sim 13 $ 
nWm$^{-2}$sr$^{-1}$ which is much larger than the estimated readout 
noise of $\sim 2.2 $ nWm$^{-2}$sr$^{-1}$. This shows that
the fluctuation of ``wide band brightness'' can solely be attributed to a celestial origin.

We applied a two-point correlation analysis to the data set after
simply subtracting the average value from ``wide band brightness''. 
The two-point correlation function is defined as 
$ C(\theta)=<\delta F(x+\theta) \delta F(x)> $, where $ \delta F(x) $ is a fluctuating 
component, $ \theta $ is the angular distance in 
degrees between two points, and $x$ represents the location of an observed point. 
Since data points are aligned along a great circle (Fig.3) and the beam pattern 
is elongated along a scan path, we pursued a one-dimensional analysis 
as a function of $ \theta $.

The top panel of Fig.17 shows our two-point correlation function on a linear scale,
in which data are binned and averaged for an  $ 0.1^{\circ} $ interval. 
Error bars are based on the scatter of $\delta F(x+\theta) \delta F(x)$, and do not
include systematic errors.
A semi-periodic component with a scale of a few degrees is found.
To confirm this feature and examine the contribution of faint stars, 
we made a similar analysis for 
the H band in the ``2MASS fluctuation data'' (middle
panel). The result shows a quite different character
from IRTS/NIRS data. The amplitude of the correlation is fairly small and no clear
spatial structure is found. This implies that the spacial fluctuation of the IRTS/NIRS
data is not stellar in origin. 
To seek any selection effect among the data points, a two-point correlation 
was also applied on a random data set at the same spacial positions but with the  
same standard deviation as ``wide band brightness''.
The result (bottom panel of Fig.17) looks like simple noise and shows no specific 
spatial structure, as expected.

In order to extract the angular scale of the fluctuations, we converted
the two-point correlation function to a power spectrum, $ P(q) $, 
where $q$ is wavenumber in units of degree$^{-1}$. We adopted a 
one-dimensional Fourier transformation of $ \theta $. 
The result is shown in Fig.18 (filled circles) where $ (q P(q))^{1/2} $ is plotted as a 
function of $q$. 
The range of the spatial frequency, $q$, is restricted to between 0.1 and 3, considering the
effective beam size (8 arcmin $\times$ 20 arcmin) and uncertainty in subtracting
zodiacal light at large angular scales. 
A clear hump at $1^{\circ} \sim 3^{\circ}$  is found.
The power spectrum for the ``2MASS fluctuation data'' is also shown by open squares in Fig.18. 
A corresponding feature at $1^{\circ} \sim 3^{\circ}$ is not found. 
100 random simulations were made, and the average values and
standard deviations (1 $\sigma$ error) are indicated by solid and dashed lines,
respectively. Fig.18 indicates that there exists statistically significant spatial fluctuation on an angular scale of $1^{\circ} \sim 3^{\circ}$ in the IRTS/NIRS data.

To confirm the characteristic angular scale of  $1^{\circ} \sim 3^{\circ}$, we divided  
the observed area into two regions, A ($ l<134^{\circ} $) and B ($ l>134^{\circ} $),
and tried to detect a similar feature in these subsets. 
These two regions are indicated in Fig.3.
 The result shown in Fig.19 is not so clear, and power spectra of the two regions 
appear fairly different. Region A has a higher angular power at low 
spatial frequency than region B.
This result indicates that the fluctuation has a broad power spectrum over 
a few-degree scale which has a distinct dependence on the sky area observed. 
To obtain any definite and decisive result, two-dimensional analysis for a
wide area is needed.

\section{Discussion}

\subsection{Comparison with previous studies}

We first review the results of EBL analysis using the COBE/DIRBE data.

Dwek and Arendt (1998) made a correlation study between the bands, 
and obtained a clear correlation between the K and L band, 
$ EBL(L)={9.9 + 0.3[EBL(K)-7.4]} \pm 2.9$ nWm$^{-2}$sr$^{-1}$. 
By adopting a value for the integrated light of galaxies at K, $ EBL(K) $, they 
reported the detection of EBL in the L band. Their result was offered
as tentative although their $ EBL(L)$ was fairly consistent with our own
values near 3.5 $\mu$m (see Table 1).

The most serious obstacle to analysis of the COBE/DIRBE data is that 
unresolved faint stars provide significant contributions to sky brightness 
due to the large beam size. Wright and Reese (2000) found
isotropic emission in the K and L bands by comparing the model Galaxy 
with the galactic distribution of the DIRBE Zodi-Subtracted data set. 
Gorjian et al. (2000) observed deep star counts
towards the COBE dark spot and subtracted the stellar contribution from
the DIRBE Zodi-Subtracted data, detecting significant isotropic
emission in the K and L bands. Recently, 2MASS data have also been used to remove the stellar
contribution from the DIRBE data (Wright 2001, Cambr\'{e}sy et al. 2001), which 
has also yielded results consistent with previous work. 

Table 2 summarizes all these results.  The isotropic emission found
by Wright and Reese (2000), Gorjian et al. (2000), and Wright (2001)
are somewhat lower than other values. This is attributable to the difference in
the IPD models used. Parenthesized values show the derived brightness when
Kelsall's IPD model is adopted. Table 2 indicates that all these observations
are consistent when corrected using a common IPD model, and demonstrates
the significant uncertainty arising from the choice of IPD model.
We deduce that the near-infrared EBL is significantly brighter
than the observed integrated light of galaxies, even allowing for all
error estimates.

\placetable{tbl-2}

\subsection{Origin of excess emission and fluctuation}

The isotropic emission detected by IRTS/NIRS is consistent with that
detected by COBE/DIRBE and is attributed to an extragalactic origin.
Furthermore, the detected emission is too bright to be explained by the 
integrated light of normal galaxies and requires events in the early Universe.

Recent observations of CMB polarization by WMAP (Kogut et al. 2003)
first determined that the reionization of the
Universe took place earlier than previously thought, back to $ z \sim17 $ or earlier.
The most probable candidates for causing the reionization are UV photons from
the first generation stars, that is, Pop.III stars. Star formation and spectral features 
of Pop.III stars have been investigated by many authors (cf. Bromm et al. 2001), 
and it is believed that massive stars are formed without metals.

Salvaterra and Ferrara (2003) tried to fit the spectrum of the excess EBL emission with
the integrated light of Pop.III stars (the solid line of Fig.13). They assumed that star 
formation of Pop.III stars terminated at $ z\sim8.8 $ to explain the spectral 
jump around 1 $\mu$m. Due to the ionized gas around stars, UV photons are
reprocessed to Lyman $\alpha$ photons which form the peak of the excess
emission around 1 $\mu$m in Fig.13.  Furthermore, they assumed a top heavy IMF
so that supernovae do not contaminate the interstellar matter with metals.
Stars responsible for the excess EBL must be heavier than 260 solar masses whose
final stage is a black hole.

The most important aspect of this paper is the significant detection of fluctuations 
in the near infrared sky. The observed fluctuations are 
fairly large ($ \sim 1/4 $ of the excess emission),
but are consistent with the result for the COBE/DIRBE data (Kashlinsky \& Odenwald 2000).
Furthermore, the two-point correlation analysis indicates that these fluctuations have 
an angular scale of a few degrees which
could be related to large scale structure. It must be noted that the 
first peak of CMB fluctuation observed by WMAP
corresponds to $1.5^{\circ}$ at the redshift of 8.8, which is very close to the
angular scale observed by IRTS/NIRS. The physical scales corresponding to
$ \sim 2^{\circ}$ are $20 Mpc$ and $200 Mpc$ for $z=8.8$ and 
the present Universe, respectively.

An important feature of the fluctuations detected by IRTS is that the spectrum
of fluctuations is very similar to that of the excess EBL. 
This could be readily understood within the model of Salvaterra and Ferrara (2003),
since the dominant contribution to the excess EBL is due to the 
nearest Pop.III stars at the terminating era. 

The amplitude and angular scale of fluctuations, however, are not
so simply understood.
Cooray et al. (2004) estimated the fluctuations assuming that density of 
Pop.III stars traces that of dark matter.  Their result shows
a broad peak at $l \sim 1000 $ which is much smaller 
than the observed angular scale of a few degrees. The estimated amplitudes of
fluctuations (zero-lag correlation) are about one order of magnitude
lower than the observed ones even for the optimistic case.  

Kashlinsky et al. (2004) made a similar analysis taking the nonlinear
effect of Pop.III star formation into account. Their result, however, still has
difficulty in explaining the large amplitudes and angular scale of  
the observed fluctuations.

In order to understand the observed fluctuations, more observational
data are needed.
In this analysis, sampling of the data points is not perfect and a one-dimensional 
analysis was applied for simplicity. Future observations with wider 
sky coverage and higher sensitivity will make two-dimensional analysis 
possible, as WMAP did. Furthermore, two-point correlation analysis
at different wavelengths will be crucial to investigate the origin of the
observed fluctuations.
 
\subsection{Connection with far infrared and submillimeter background}

In the far infrared and submillimeter region, detections of the EBL are also reported
based on the COBE/DIRBE data (Hauser et al. 1998,  Lagache et al. 2000,  
Finkbeiner et al. 2000, Matsuhara et al. 2000) and COBE/FIRAS data 
(Puget et al. 1996, Fixsen et al. 1998).
These EBLs should be compared with the integrated light from point sources.

 At submillimeter wavelengths,  deep surveys using lensing effects 
with the SCUBA instrument on the JCMT measurements
have shown that SCUBA sources at 850 $\mu$m
can explain most of the submillimeter background (Blain et al 1999). 
Counterparts to SCUBA sources are thought to be ultraluminous galaxies
with a median redshift $z \sim 2.4$ (Smail et al. 2002, Chapman et al. 2004).

In the far-infrared region, ISO surveys 
had been the sole available deep observations. 
The detection limits of the ISO surveys, however, 
were not deep enough to estimate the contribution
of point sources to the EBL (Matsuhara et al. 2000). 
Recent MIPS/SST observations, however, provided a several times 
deeper survey than ISO (Dole et al. 2004), but could resolve only 23 \% 
and 7 \% of the cosmic far-infrared background at 70 $\mu$m and 160 $\mu$m, respectively. 
Much deeper surveys with advanced space 
telescopes such as Herschel Space Telescope (Pilbratt 2004), 
SPICA (Matsumoto 2004), etc. are required to understand the origin of far-infrared EBL.
 
In any case, the origin of the excess emission in the far-infrared region 
must be thermal emission from the dust. Since Pop.III stars are formed from
Hydrogen and Helium atoms without metals, the far-infrared EBL must
originate in a period later than the Pop.III era. The far infrared
EBL could be a key to understanding the star formation history after the Pop.III era.

\subsection{Future observations} 

The near infrared background is a unique observational tool to investigate the Pop.III era.
From the observational point of view, there are three important aspects that
are expected of future observations.
The first is to confirm the spectral jump around 1 $\mu$m. This is important
to investigate the termination era of Pop.III star formation. The second is to 
measure the fluctuations at the different angular scales and wavelengths. This
will provide valuable information to understand the
evolution of Pop.III stars and large scale structure. The third is absolute
measurement, free from the uncertainty in IPD model.

It is not so easy to achieve these three criteria, since space observations
are absolutely necessary.
The Japanese ASTRO-F mission that will be launched in 2005 
will be a powerful tool 
in near infrared background observations (Murakami 2004). Its Infrared Camera (IRC) 
will provide images of 10 arcmin over pixel scales of 1.5 arcsec for the K and L bands.
A large area survey towards the north ecliptic pole is being planned (Pearson et al. 2004)
and the observation of 
fluctuations from a few arcsec to several degrees will be attained with excellent quality 
(Cooray et al. 2004). Low resolution spectroscopy at wavelengths longer than
1.8 $\mu$m will also provide additional spectral information.

The shortest wavelength of the IRC is 2 $\mu$m, and no satellite mission
is scheduled to observe the background emission around 1 $\mu$m.
Owing to the progress of detector arrays, even sounding rocket experiments with 
short observing times can provide data of high quality. 
The rocket experiment, CIBER (Cosmic Infrared Background ExpeRiment) has been 
approved by NASA. CIBER is a US-Japan-Korea joint mission (PI: J. Bock of JPL),
and has two objectives.  One is to measure the 
spectrum of the sky around 1 $\mu$m, and
another is to measure fluctuations of sky at the I and H bands with a wide field camera. 
CIBER is a complementary experiment to the ASTRO-F mission that can be 
realized in a relatively short period at low cost.

Absolute measurement is the most difficult aspect. Since it is difficult to improve the
IPD model, the best technique will be to measure sky brightness outside the zodiacal
cloud. ISAS/JAXA is now examining a solar sail mission to Jupiter.
The near infrared instrument for the measurement of background radiation
is proposed as one of the onboard instruments (Matsuura et al. 2004). At the orbit
of Jupiter, the surface brightness of the zodiacal light towards the ecliptic pole is estimated
to be $ \sim 1/20 $ of that at the Earth's orbit, and $1/4$ of the observed near infrared EBL. 
A solar sail mission towards Jupiter will definitely render accurate absolute brightness
and fluctuation measurements of the near infrared EBL.

\section{Conclusion}  

The IRTS/NIRS has provided high quality data with which to seek the near infrared
extragalactic background light (EBL) with an 8 arcmin $\times$ 20 arcmin beam, across a 
wavelength range from 1.4 $\mu$m to 4.0 $\mu$m, with spectral resolution of
0.13 $\mu$m. The contribution of the foreground emission  
(zodiacal component and integrated light of faint stars) was subtracted by 
modeling, and significant isotropic emission was detected that we attribute 
to an extragalactic origin. 
Its spectrum is star-like but shows a spectral discontinuity with the optical EBL. 

The observed isotropic emission is significantly brighter than that of the integrated light 
of faint galaxies. Our results are consistent with the results of 
independent analyses of star-subtracted DIRBE data, although significant
uncertainty stems from the use of different IPD models. 

The spectrum and brightness of the observed excess EBL emission could
be explained by the
redshifted UV radiation from the first stars (Pop.III stars) that caused the reionization
of the Universe. WMAP observations of
the CMB polarization has indicated that reionization occurred at $ z\sim 17 $ 
or earlier, while the spectral jump around 1 ${\mu}$m of observed excess EBL
suggests that the Pop.III star formation terminated at $ z \sim 9 $.

Excess fluctuations of the sky brightness were detected at a level $ \sim 1/4 $ of 
the excess emission. A two-point correlation analysis indicates that the 
excess fluctuations have an angular scale of a few degrees. 
The observed fluctuations, however, are considerably larger than the 
theoretical predictions for the Pop.III stars.

\section{ACKNOWLEDGMENTS}  

The authors thank IRTS members for their encouragement, and 
Dr. Totani and Prof. Takahashi for inspiring discussions.  
Thanks are also due to Drs. Ferrara and Salvaterra for their courtesy in using their
theoretical model. We are also grateful to Ms. Kimura for 
her support in acquiring the 2MASS data. 

MC thanks NASA for supporting his participation
in IRTS/NIRS work through Co-operative Agreement NCC2-1125 between Berkeley
and NASA-Ames Research Center.

This publication makes use of data products from the Two Micron All Sky Survey, 
which is a joint project of the University of Massachusetts and the Infrared 
Processing and Analysis Center/California Institute of Technology, funded by 
the National Aeronautics and Space Administration and the National Science Foundation.

\clearpage


\begin{figure} 
\plotone{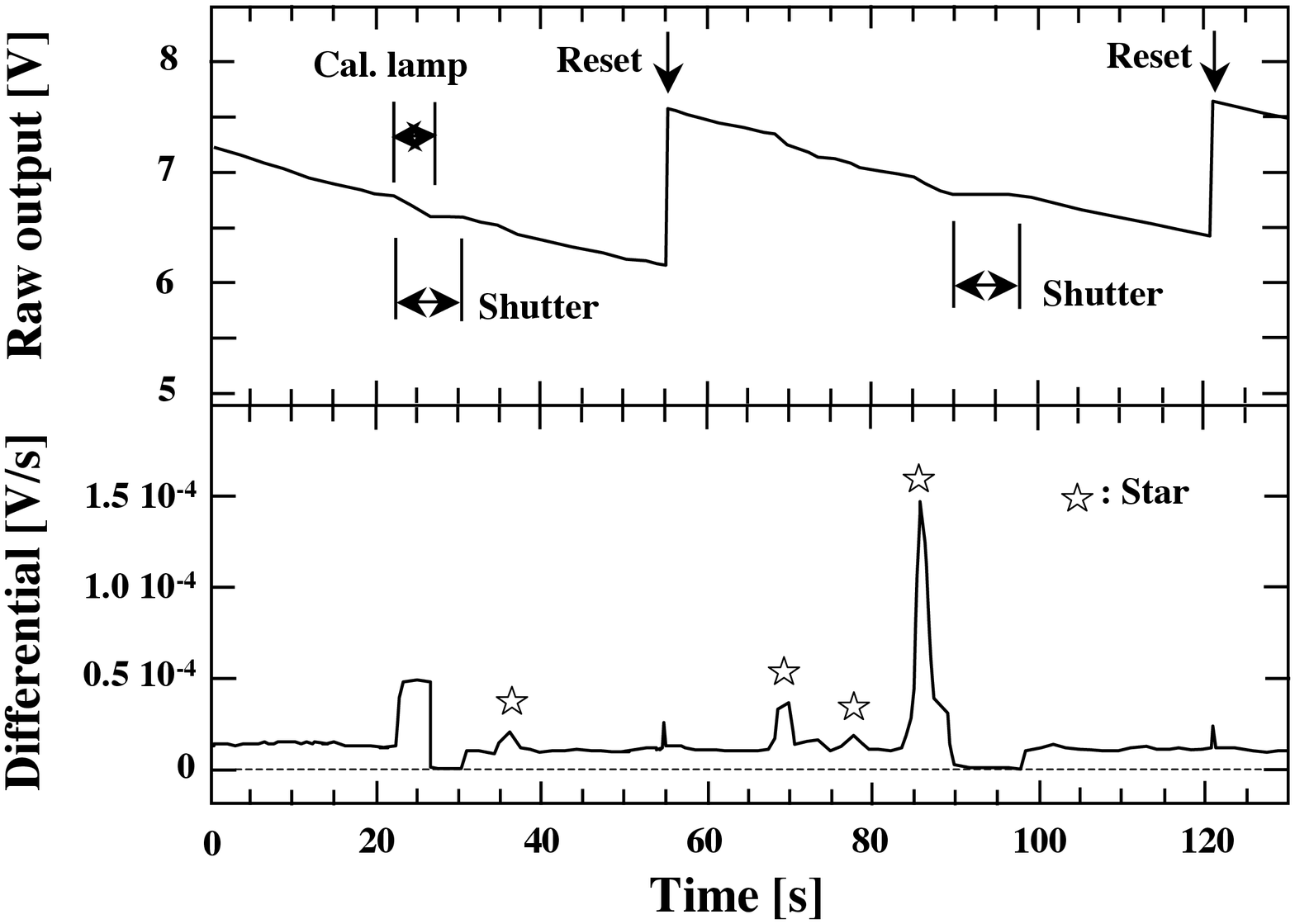}
\caption{The raw output signal of a typical wavelength band is illustrated. 
	Upper panel shows the ramp curve, while the lower panel indicates the differentiated
	signal in which incident stars are shown by 5-pointed symbols. 
      	 \label{fig1}}
\end{figure}

\clearpage

\begin{figure}
\plotone{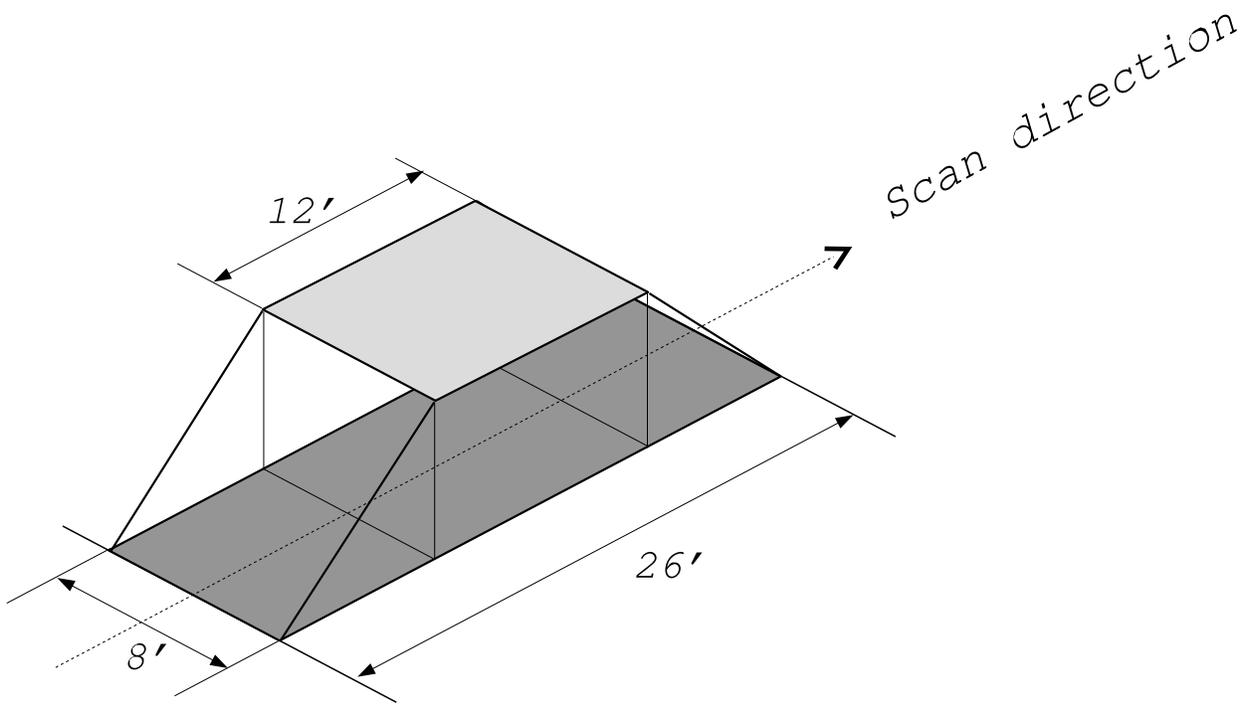}
\caption{The beam pattern for 5-sec integration data. The trapezoidal
	 pattern elongated along the scan direction is shown.  
        \label{fig2}}
\end{figure}

\clearpage

\begin{figure}
\plotone{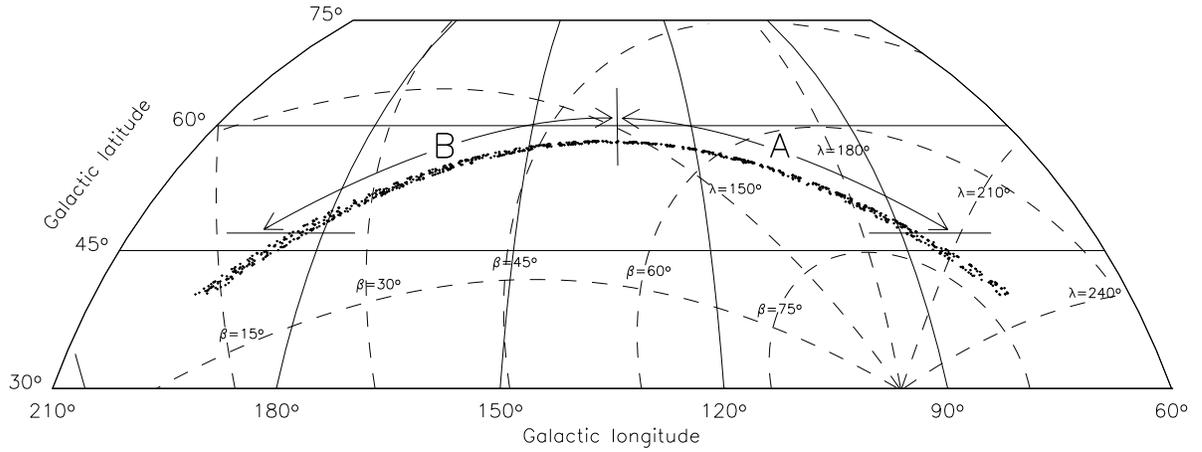}
\caption{The positions of the spectra acquired in galactic coordinates.
	Dashed lines indicate ecliptic coordinates. The Sun lay close to the
	Spring equinox, and solar elongation angles were $\sim 90^{\circ}$ 
	during the observations. The region designated by arrows is  
	used for the fluctuation analysis (see section 4).
        \label{fig3}}
\end{figure}

\clearpage

\begin{figure}
\plotone{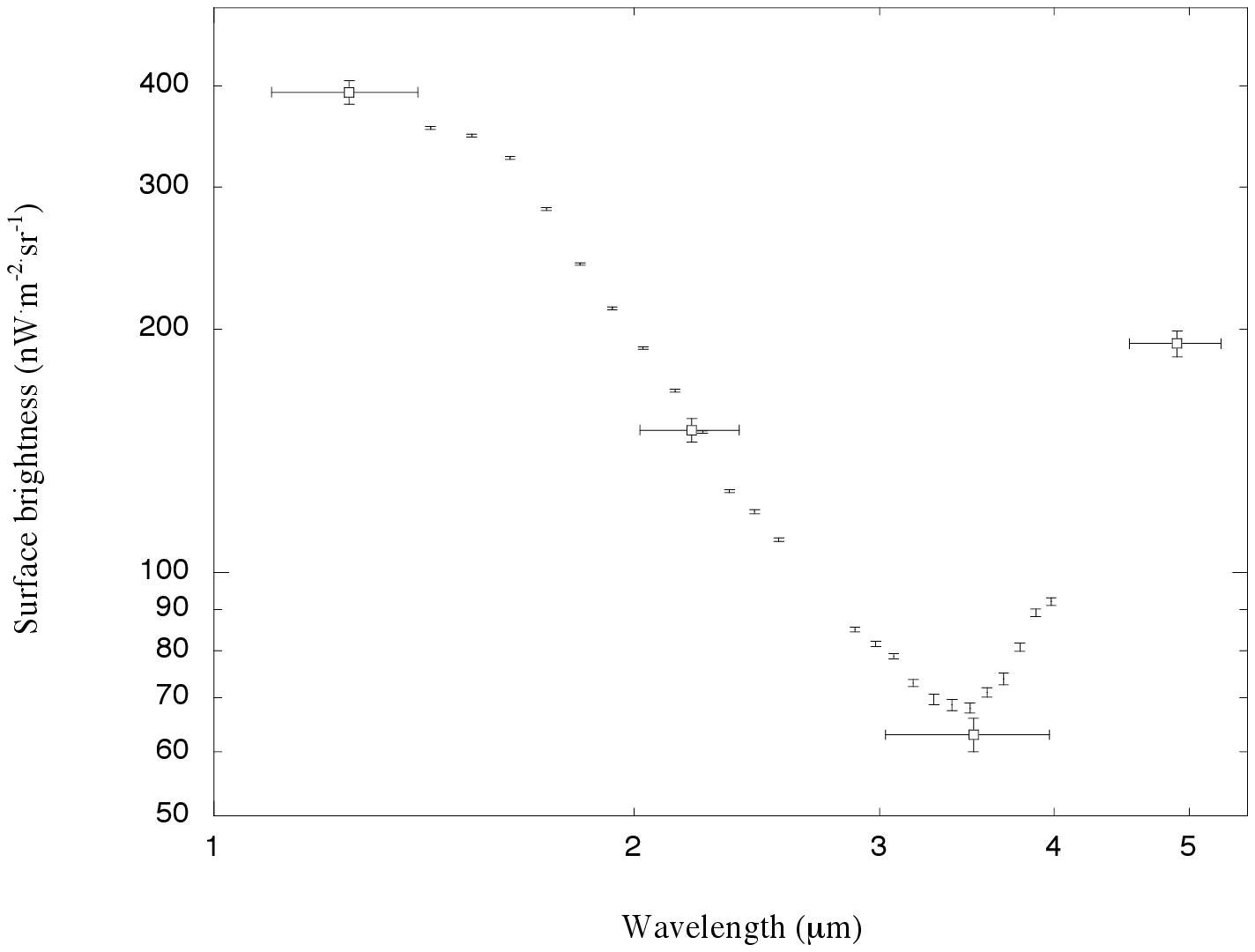}
\caption{The averaged spectrum of the sky at high ecliptic latitude 
        ($ \beta>70^{\circ}$). Vertical bars indicate the uncertainties which 
        are dominated by systematic errors in calibration. The open squares indicate 
        the DIRBE darkest sky for the J, K, L and M bands (Hauser et al. 1998). 
        \label{fig4}}
\end{figure}

\clearpage

\begin{figure}
\plotone{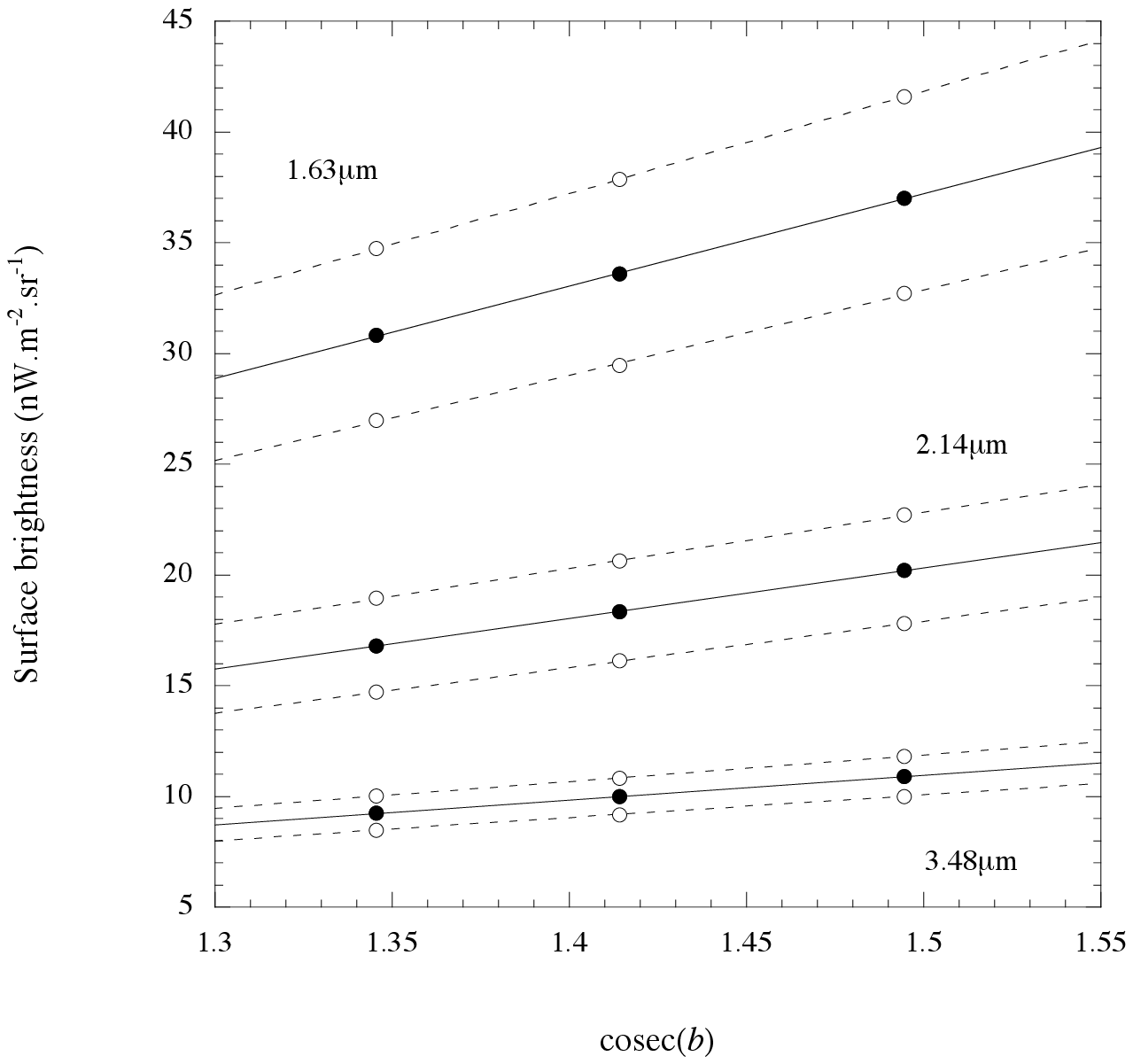}
\caption{ The integrated light of faint stars at $ b=42^{\circ}, 45^{\circ}, 48^{\circ} $
	based on the SKY model is shown as a function of $ cosec(b) $ by filled circles.	
	Open circles indicate upper and lower limits due to errors of the cut-off
	magnitudes ($\pm0.5$). The solid and dashed lines are drawn by linear
	fitting.
	From the top, the results for 1.63, 2.14 and 3.48 $\mu$m 
	bands are shown, respectively.
	\label{fig5}}
\end{figure}

\clearpage

\begin{figure}
\plotone{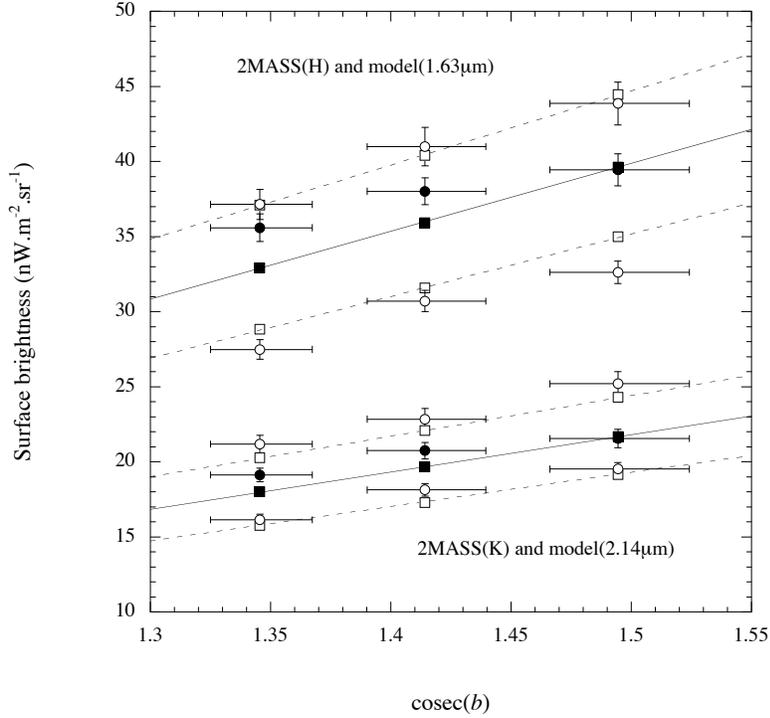}

\vspace{-3cm}

\caption{The integrated light of faint stars at 1.63 and 2.14 $\mu$m
	based on the SKY model (squares) is compared with the integrated light of 
	2MASS stars at the H and K bands (circles). Filled squares and circles
	show the results for the nominal cut-off magnitude at 
	$ b=42^{\circ}, 45^{\circ}, 48^{\circ} $, while open squares
	and circles indicate the results for cut-off magnitudes $\pm0.5$ for the same
	galactic latitudes. 
	Solid and dashed lines are drawn adopting linear fitting of the model.
	Model calculation and 2MASS data acquisition were made for a flat
	top beam,
	\label{fig6}}
\end{figure}

\clearpage

\begin{figure}
\plotone{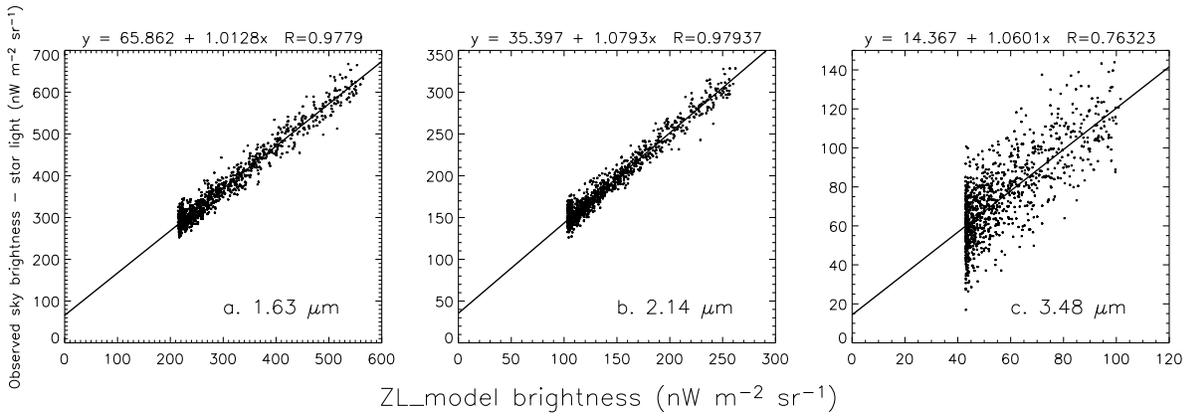}
\caption{ Correlation diagrams between the star-subtracted brightness and the 
	modeled zodiacal component at (a) 1.63 $\mu$m, (b) 2.14 $\mu$m 
     	and (c) 3.48 $\mu$m, respectively. 
	\label{fig7}}
\end{figure}

\clearpage

\begin{figure}
\plotone{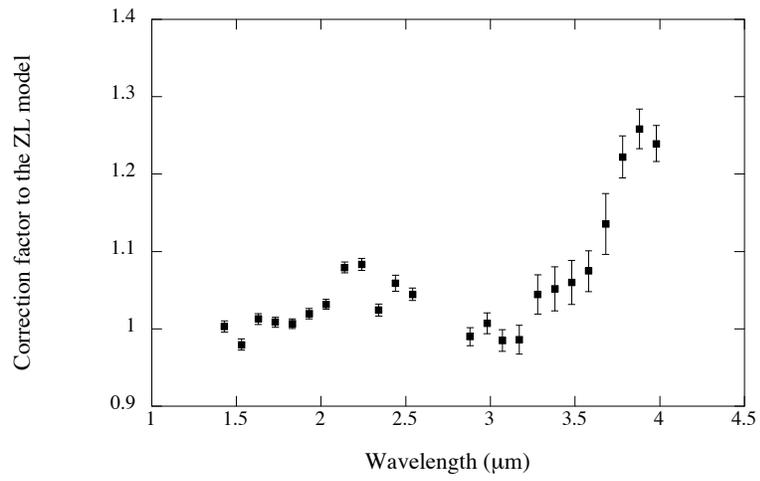}
\caption{Correction factor to the model zodiacal component is shown 
	for the 24 NIRS wavelength bands.	
	\label{fig8}}
\end{figure}

\clearpage

\begin{figure}
\begin{center}
\includegraphics{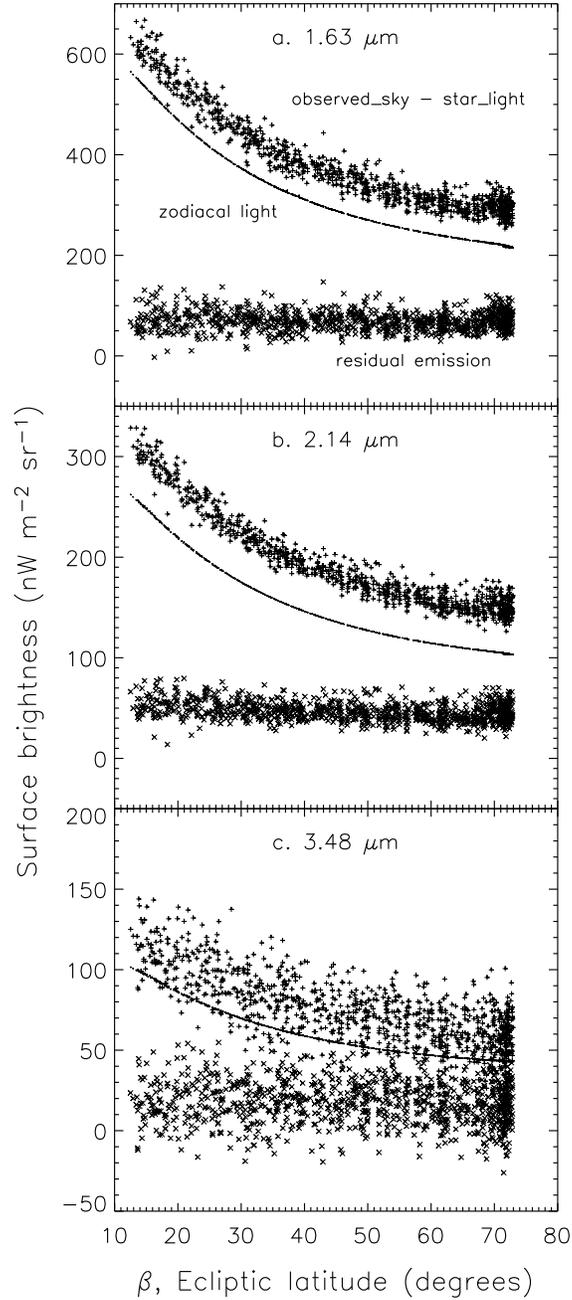}
\end{center}


\caption{ The ecliptic latitude dependences of (a) 1.63 $\mu$m, 
         (b) 2.14 $\mu$m and (c) 3.48 $\mu$m are shown. The pluses, 
	 dots (look like solid lines), and crosses indicate the
         star-subtracted brightness, the modeled zodiacal components, 
         and the residual brightness after subtracting the 
         zodiacal components from the star-subtracted brightness, respectively.
         The model zodiacal components are corrected to the original model (Kelsall et al. 1998)
	 by the correction factors in Fig.8.
	\label{fig9}}
\end{figure}

\clearpage

\begin{figure}
\plotone{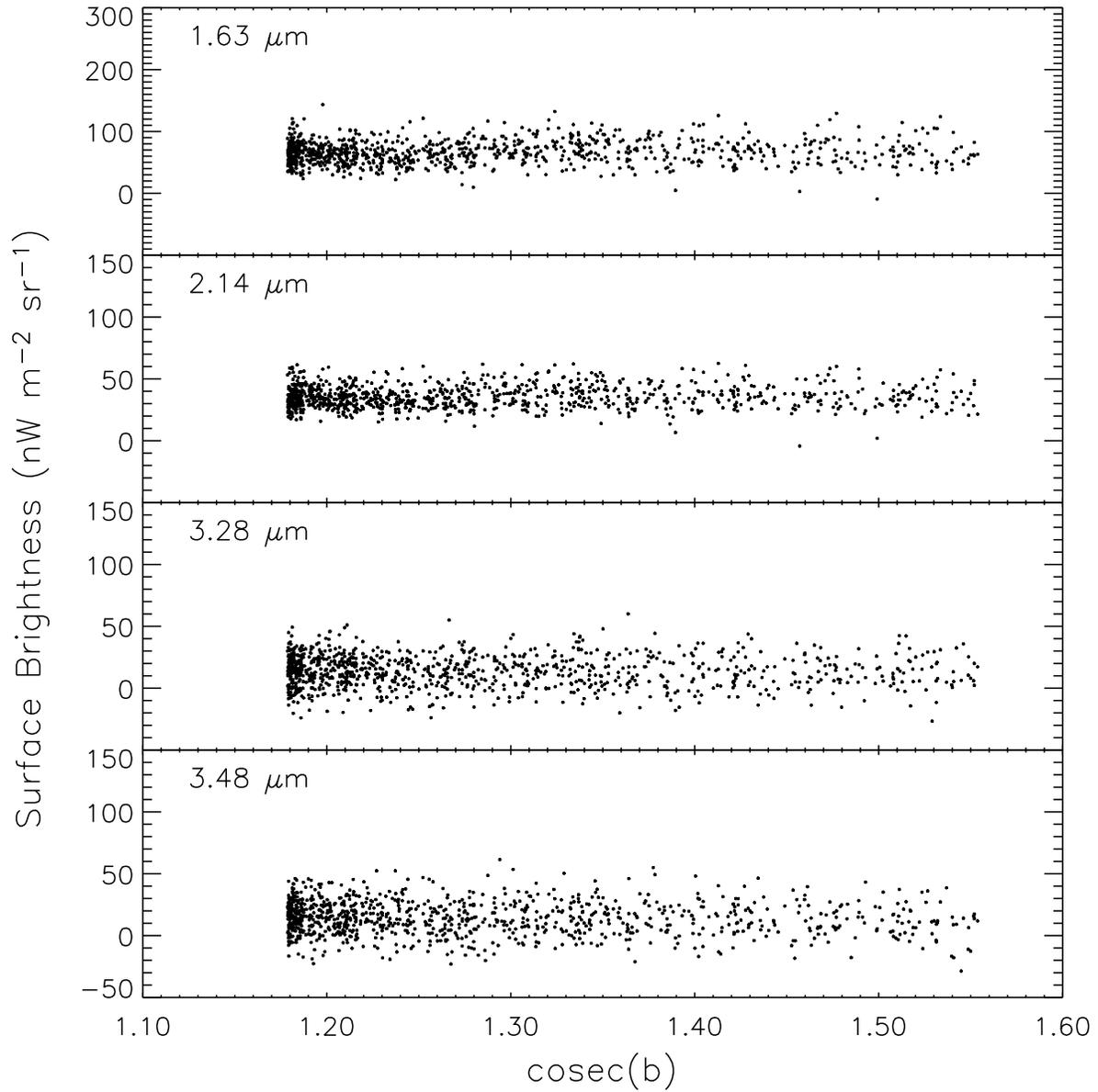}
\caption{ The galactic latitude dependences of (a) 1.63 $\mu$m, 
         (b) 2.14 $\mu$m and (c) 3.28 (d) 3.48 $\mu$m for the 
	residual emission are shown as a function of $ cosec(b) $. 
	\label{fig10}}
\end{figure}

\clearpage

\begin{figure}
\plotone{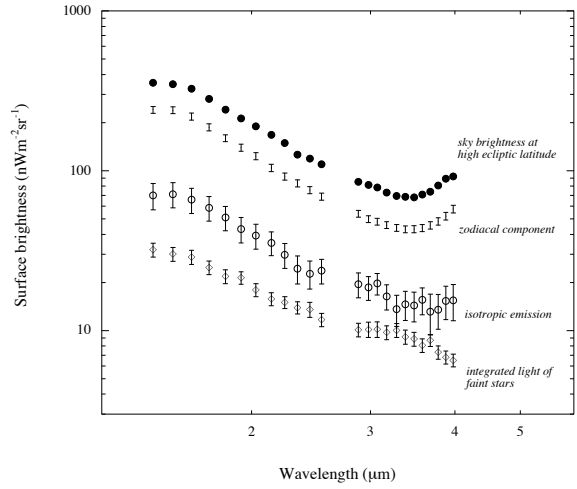}
\caption{ The breakdown of sky brightness at high ecliptic latitude
        is shown. From the top, the spectra of the dark sky (filled 
        circles, as in Fig.4),  the zodiacal component (bars), the  
        isotropic emission (open circles) and the integrated light of faint stars 
        (open diamonds) are indicated, respectively.
	\label{fig11}}
\end{figure}

\clearpage

\begin{figure}
\plotone{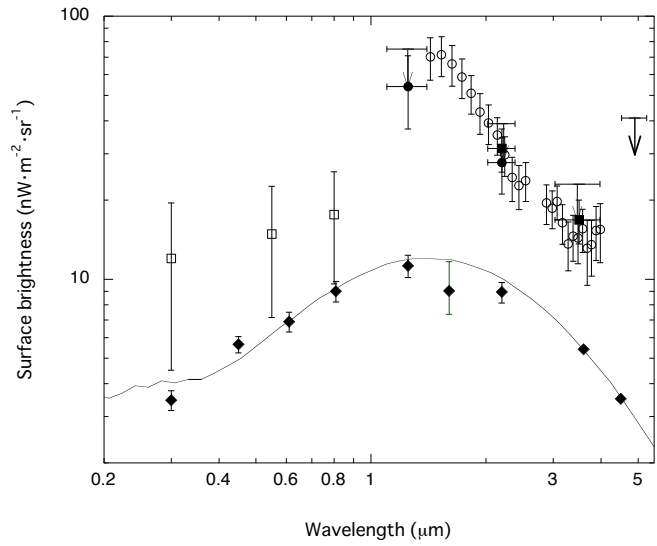}

\vspace{0cm}

\caption{The spectrum of the observed isotropic emission is shown by open circles, 
        while upper limits from COBE/DIRBE data (Hauser et al. 1998) are
        shown by arrows. The optical EBL is presented
	by the open squares (Bernstein et al. 2002).
	The EBL obtained from the star-subtracted COBE/DIRBE data is shown  
	by the filled squares (Wright and Reese 2000)
	and the filled circles (Cambr\'{e}sy et al. 2001).
    	The data by Wright and Reese (2000) have been modified to use
	Kelsall's IPD model.
        The filled diamonds represent the integrated light of galaxies compiled 
        by Madau and Pozzetti (2000) for the H band, by Fazio et al (2004) for the
	3.6 and 4.5 $\mu$m bands, and by Totani et al. (2001) for other bands. 
	The solid line shows the theoretical model of the EBL by Totani and Yoshii (2000). 
	\label{fig12}}
\end{figure}

\clearpage

\begin{figure}
\plotone{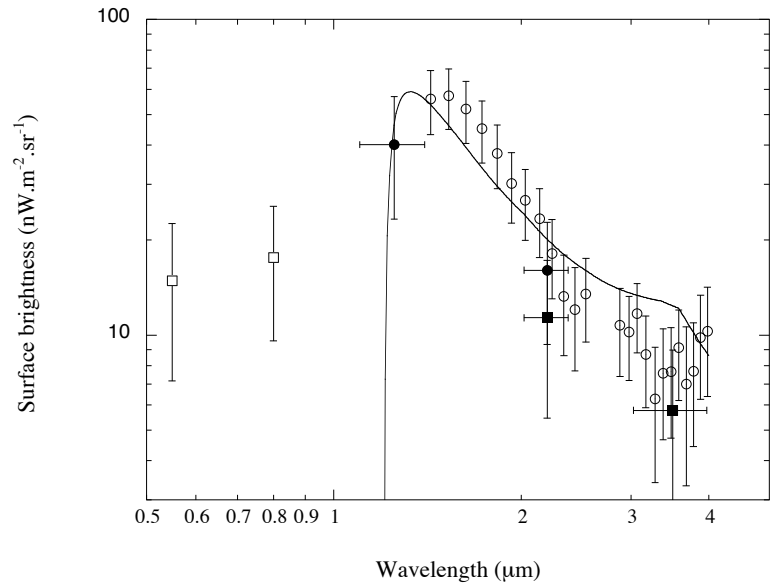}

\vspace{-3cm}

\caption{ The spectrum of the excess brightness over the integrated
	light of galaxies is shown by the open circles (IRTS/NIRS),
	the filled squares (Wright and Reese 2000) and the 
	filled circles (Cambr\'{e}sy et al. 2001). 
	The optical EBL is represented by the open squares (Bernstein et al. 2002).
	The solid line indicates the model brightness by Salvaterra and Ferrara (2003).
	\label{fig13}}
\end{figure}

\clearpage

\begin{figure}
\plotone{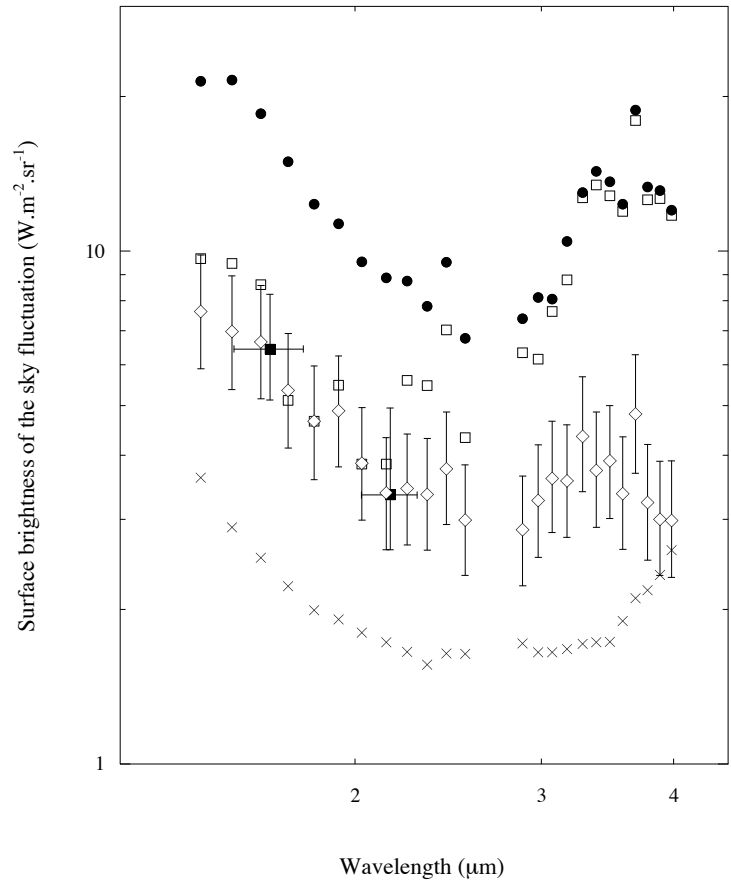}

\vspace{-2cm}

\caption{The spectrum of fluctuations observed by IRTS/NIRS is shown by 
        the filled circles. The open squares, the open diamonds and the crosses represent
	fluctuations due to read out noise, star light, and photon noise, respectively.
	The filled squares indicate fluctuations due to 2MASS stars for the H and
	K bands.
         \label{Fig14}}
\end{figure}

\clearpage

\begin{figure}
\plotone{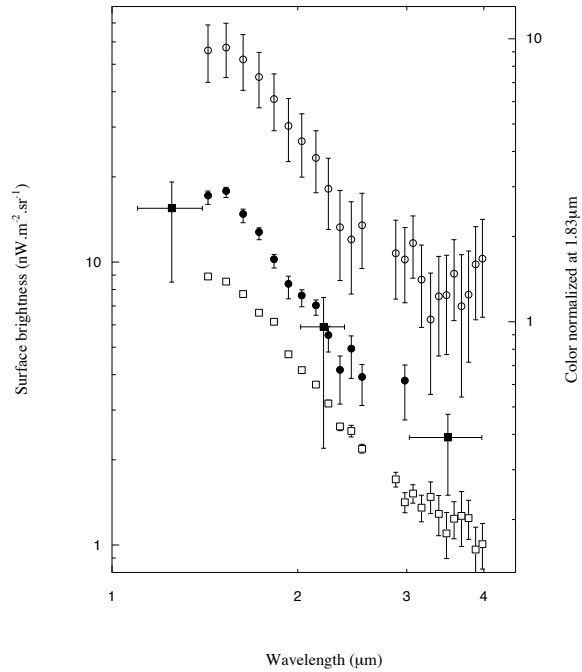}
\caption{ The spectrum of the excess fluctuation (filled circles) is compared with 
	the excess brightness of the isotropic emission (open circles). The spectrum 
	of the fluctuations obtained from the correlation analysis
	is also shown by the open squares with arbitrary units on the right-side ordinate. 
	The filled squares are fluctuations obtained from the DIRBE data 
	by Kashlinsky and Odenwald (2000).
	\label{Fig15}}
\end{figure}

\clearpage

\begin{figure}
\begin{center}
\includegraphics{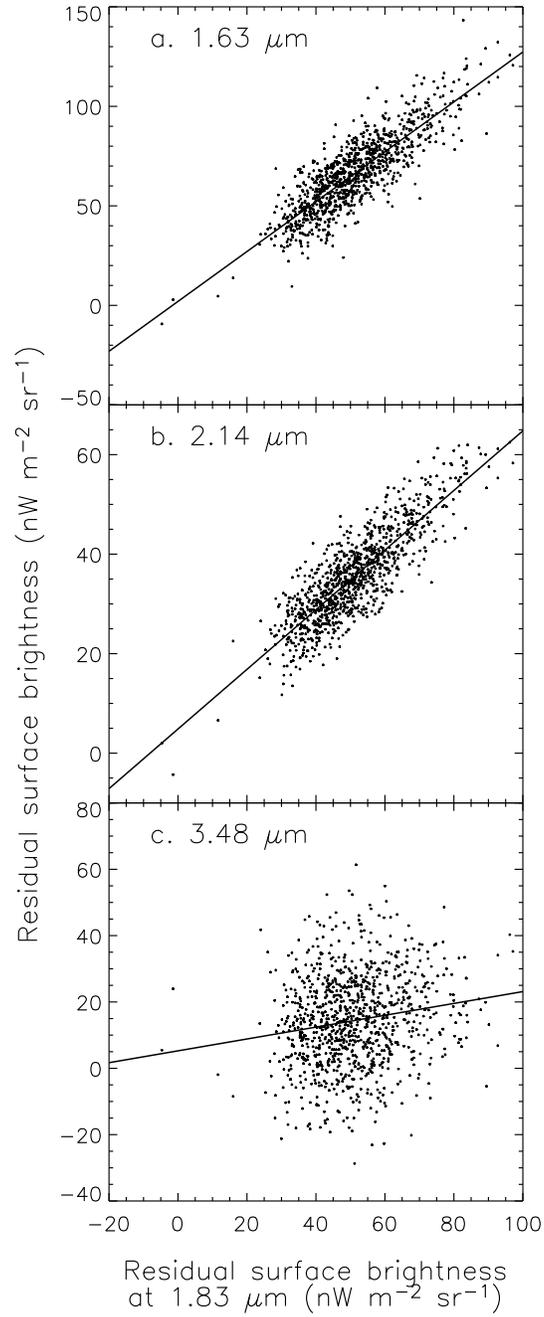}
\end{center}
\caption{Correlation diagrams between the residual emission at
	a) 1.63 $\mu$m, b) 2.14 $\mu$m and  c) 3.48 $\mu$m  
        and that at 1.83 $\mu$m. The solid lines indicate the resultant linear fitting. 
	\label{Fig16}}
\end{figure}

\clearpage

\begin{figure} 
\plotone{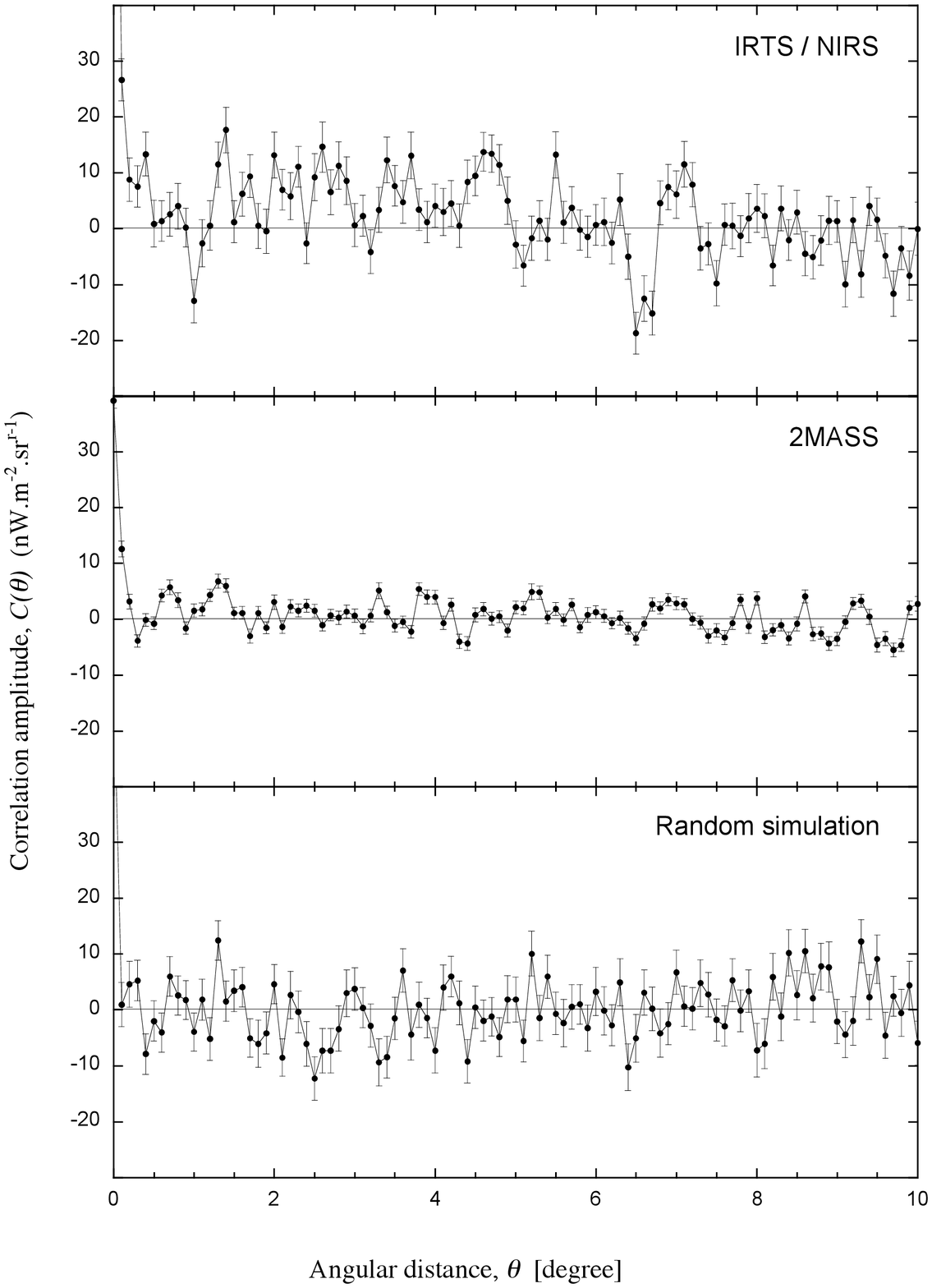}
\caption{ A two-point correlation function is shown as a function of
        	angular distance $\theta$. From the top, the results for
	IRTS/NIRS, the 2MASS stars, and the random simulation, are shown, respectively.
	\label{Fig17}}
\end{figure}

\clearpage

\begin{figure}
\plotone{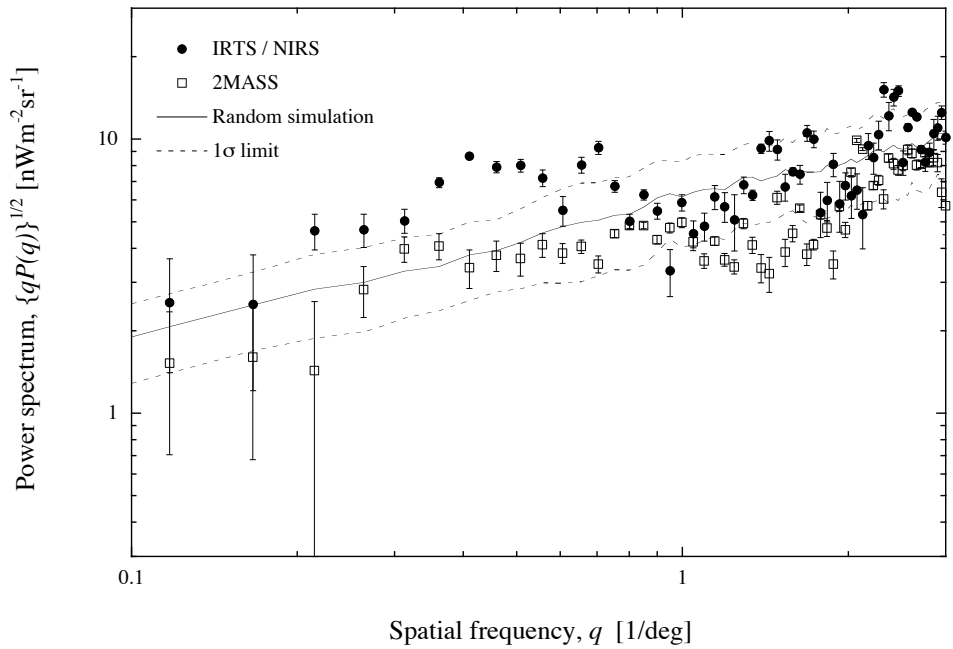}
\caption{ A power spectrum 
        $ (q P(q))^{1/2} $ is shown as a function of spatial frequency, $q$ in units of degree$^{-1}$. 
	The filled circles and the open squares indicate the results 
	for the ``wide band brightness'' of the
	IRTS/NIRS data, and for the 2MASS data, respectively.
	The solid line shows the mean value of the random simulations, 
	while dotted lines indicate the $ \pm1 \sigma $ error range. 
	\label{Fig18}} 
\end{figure}

\begin{figure}
\plotone{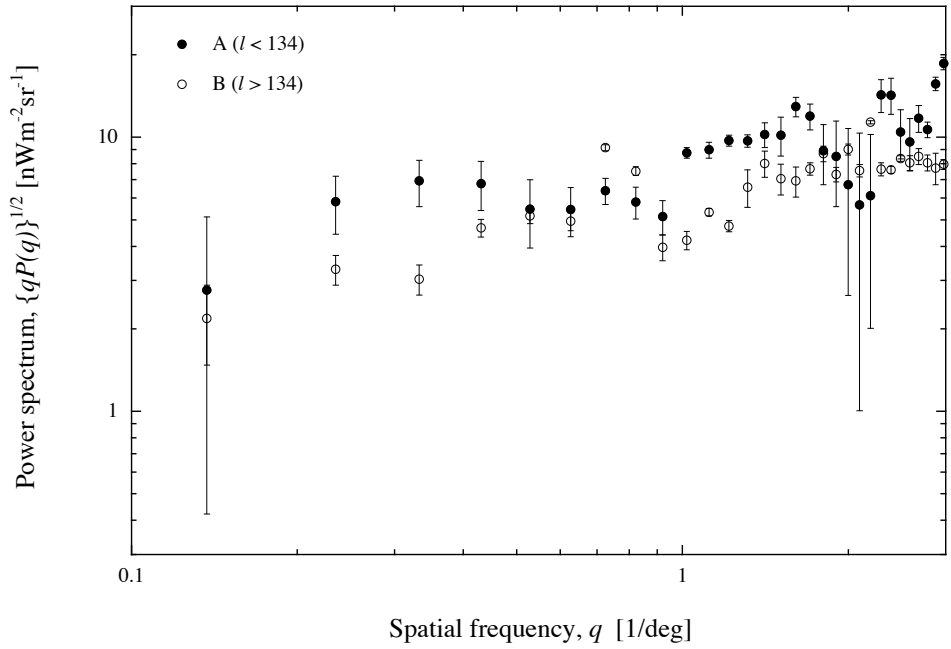}
\caption{ Power spectra for two subsets are
       	shown as a function of spatial frequency, $q$ in units of degree$^{-1}$. 
	The filled circles and the open squares indicate the results for
	region A($ l<134^{\circ} $) and B($ l>134^{\circ} $), respectively.
	\label{Fig19}} 
\end{figure}

\clearpage

\begin{deluxetable}{ccccccccc}
\tabletypesize{\small}
\tablewidth{0pc}
\tablecolumns{9}
\tablecaption{Surface brightness of residual isotropic emission and its errors
in units of nWm$^{-2}$sr$^{-1}$ \label{tbl-1}}
\tablehead{
\colhead{}&
\colhead{}&
\colhead{}&
\multicolumn{6}{c}{Errors}\\
\cline{4-9}\\
\colhead{Wavelength} &
\colhead{Residual} &
\colhead{}&
\colhead{Fitting} &
\multicolumn{2}{c}{Error of Starlight} &
\colhead{Error of} &
\colhead{Systematic} &
\colhead{Total}\\
\cline{5-6}
\colhead{($\mu$m)}&
\colhead{}&
\colhead{}&
\colhead{error}&
\colhead{Cut-off Mag} & 
\colhead{Model} &
\colhead{ZL Model} &
\colhead{error} &
\colhead{}
}
\startdata 
3.98 & 15.5 && 1.73 & 0.47 & 0.58 & 3.17 & 1.37 & 3.9 \\
3.88 & 15.3 && 1.75 & 0.48 & 0.61 & 2.83 & 1.15 & 3.6 \\
3.78 & 13.5 && 1.74 & 0.51 & 0.66 & 2.57 & 0.83 & 3.3 \\
3.68 & 13.1 && 2.40 & 0.66 & 0.78 & 2.38 & 1.22 & 3.7 \\
3.58 & 15.5 && 1.56 & 0.58 & 0.73 & 2.24 & 0.79 & 3.0 \\
3.48 & 14.4 && 1.67 & 0.60 & 0.80 & 2.17 & 0.88 & 3.0 \\
3.38 & 14.6 && 1.69 & 0.67 & 0.82 & 2.13 & 0.77 & 3.0 \\
3.28 & 13.6 && 1.54 & 0.71 & 0.90 & 2.17 & 0.81 & 3.0 \\
3.17 & 16.4 && 1.19 & 0.70 & 0.88 & 2.25 & 0.98 & 2.9 \\
3.07 & 19.7 && 0.94 & 0.93 & 0.92 & 2.35 & 1.06 & 3.0 \\
2.98 & 18.6 && 0.94 & 0.77 & 0.92 & 2.44 & 1.35 & 3.2 \\
2.88 & 19.5 && 0.88 & 0.91 & 0.92 & 2.67 & 1.56 & 3.5 \\
2.54 & 23.7 && 0.79 & 1.07 & 1.07 & 3.55 & 1.31 & 4.2 \\
2.44 & 22.7 && 1.11 & 1.06 & 1.23 & 3.91 & 1.04 & 4.5 \\
2.34 & 24.4 && 0.92 & 1.28 & 1.27 & 4.28 & 0.92 & 4.8 \\
2.24 & 29.7 && 1.01 & 1.38 & 1.37 & 4.77 & 0.71 & 5.3 \\
2.14 & 35.4 && 1.03 & 1.66 & 1.46 & 5.38 & 0.77 & 6.0 \\
2.03 & 39.2 && 1.12 & 2.08 & 1.67 & 6.28 & 0.74 & 7.0 \\
1.93 & 43.2 && 1.34 & 2.32 & 1.98 & 7.08 & 0.89 & 7.9 \\
1.83 & 51.0 && 1.45 & 2.56 & 2.04 & 8.01 & 0.88 & 8.8 \\
1.73 & 58.7 && 1.74 & 2.85 & 2.31 & 9.38 & 1.03 & 10.3 \\
1.63 & 65.9 && 2.14 & 3.24 & 2.68 & 10.8 & 1.54 & 11.9 \\
1.53 & 71.3 && 2.45 & 3.26 & 2.80 & 11.6 & 2.11 & 12.8 \\
1.43 & 70.1 && 2.48 & 3.72 & 2.98 & 11.8 & 2.53 & 13.2 \\
\enddata 
\end{deluxetable} 

\clearpage

\begin{table} 
\caption{Results for the near infrared EBL based on the COBE/DIRBE data
         are summarized. Numbers in parentheses are values modified to use
         Kelsall's IPD model. Units are nWm$^{-2}$sr$^{-1}$. \label{tbl-2}}
\begin{center}
\begin{tabular}{lccc}
	\hline
	\hline
	Authors & J & K & L \\
	\hline
	Dwek \& Arendt (1998) & -- & -- & $9.9\pm 2.9$ \\
	Gorjian et al. (2000) & -- & $22.4\pm 6.0$ & $11.0\pm 3.3$\\
	& -- & $(30.7\pm 6.0)$ & $(15.4\pm 3.3)$ \\
	Wright and Reese (2000) & -- & $23.1\pm 5.9$ & $12.4\pm 3.2$\\
	& -- & $(31.4\pm 5.9)$ & $(16.8\pm 3.2)$\\
	Wright (2001) & $28.9\pm 16.3$ & $20.2\pm 6.3$ & --\\
	& $(61.9\pm 16.3)$ & $(28.5\pm 6.3)$ & --\\
	Cambr\'{e}sy et al. (2001) & $54.0\pm 16.8$ & $27.8\pm 6.7$ & -- \\
	\hline
\end{tabular}
\end{center}
\end{table}

\end{document}